# Enhanced critical current properties in $Ba_{0.6}K_{0.4+x}Fe_2As_2$ superconductor by over-doping of potassium

**Chunlei Wang, Lei Wang, Zhaoshun Gao, Chao Yao, Dongliang Wang, Yanpeng Qi, Xianping Zhang, and Yanwei Ma[1]**

Key Laboratory of Applied Superconductivity, Institute of Electrical Engineering, Chinese Academy of Sciences, P.O. box 2703, Beijing 100190, People's Republic of China

Phase-pure polycrystalline $Ba_{0.6}K_{0.4+x}Fe_2As_2$ with $0 \leq x \leq 0.1$ were prepared using a one-step solid-state reaction method. We found that over-doping of potassium can improve critical current density ($J_c$). High-field $J_c$ for samples with $x = 0.1$ is three times higher than that for samples with $x = 0$. Over-doping of K has minimal effect on the critical transition temperature ($T_c$). Less than 0.5 $K$ degradations in $T_c$ was measured for samples with $x = 0.1$. TEM revealed high concentration of dislocations in samples with $x = 0.1$, resulting in enhanced flux pining. Further analyses on magnetization loops for powder samples confirm that K over-doping can promote intra-grain $J_c$. Our results indicate that slight excess of K in $Ba_{0.6}K_{0.4}Fe_2As_2$ samples is beneficial to high-field applications.

[1] Author to whom any correspondence should be addressed.
E-mail: ywma@mail.iee.ac.cn

The discovery of high-$T_c$ superconductivity in iron-based LaO$_{1-x}$F$_x$FeAs superconductor by Kamihara et al [1] has triggered renewed research interests in superconducting science and technology [2-4]. Ba$_{0.6}$K$_{0.4}$Fe$_2$As$_2$ was reported to exhibit superconductivity at 38 $K$ in 2008 [5]. Shortly after, superconductivity was reported in SrFe$_2$As$_2$, CaFe$_2$As$_2$ and EuFe$_2$As$_2$ by appropriate substitution or under pressure [6-8]. The AFe$_2$As$_2$ (where A = Ba, Sr, Ca, Eu) compounds are built up with identical FeAs layers separated by A, so there are double Fe-As layers in a unit cell. It is widely accepted that these materials represent the second class of high-$T_c$ superconductors after the discovery of cuprates in 1986.

The K doped BaFe$_2$As$_2$ exhibits moderately higher critical transition temperature (about 38 $K$), high critical current density $J_c$, low anisotropy, and high critical fields [9-10]. When compared with that of the RE-1111 series, it requires lower synthesis temperature (800~900$^{\circ}$C). Especially, it has been reported that single crystal Ba$_{1-x}$K$_x$Fe$_2$As$_2$ show very strong intrinsic pining and high $J_c$ [11-13]. All these are desirable for practical applications.

Chemical addition could play an important role in enhancing superconducting properties, by promoting the crystallization of the superconducting phase, catalyzing the intergranular coupling of the superconducting grains or introducing pinning centers [14-16]. For example, Fang et al reported that adding 5 wt % of excess Mg in a Cu-clad wire resulted in a significant improvement in the critical current density [17], and others found that B-rich and Mg-rich MgB$_2$ have different $J_c$-$B$ behaviors [18-19]. Like Mg in MgB$_2$, element K is also very active and has low melting point. As a result, burning loss of K is unavoidable, usually leading to K-deficient in Ba$_{1-x}$K$_x$Fe$_2$As$_2$ superconductor. Thus, appropriate over-doping of K in Ba$_{1-x}$K$_x$Fe$_2$As$_2$ may have positive effect on the superconducting properties. However, there are no related studies about the influence of potassium over-doping on Sr/Ba-122 system until now. In this paper, we report the effect of over-doping of K on the microstructure and superconducting properties of polycrystalline Ba$_{0.6}$K$_{0.4}$Fe$_2$As$_2$. An enhancement of $J_c$ in pnictide bulks can be achieved by over-doping of K.

Polycrystalline Ba$_{0.6}$K$_{0.4+x}$Fe$_2$As$_2$ (x = 0, 0.02, 0.04, 0.06, 0.08, 0.1) bulk samples were synthesized through a one-step method developed in our group [20]. Stoichiometric amount of Ba filings, Fe powder, As and K pieces, were ball milled in Ar atmosphere for ~ 6 hours to achieve better uniformity. In order to compensate the loss of As at high temperatures, extra 5 wt.% As was added. The powder was filled into a Ta tube with 8 mm outer diameter and 1 mm wall thickness. After packing, the tube was rotary swaged with final outside diameter of 5.9 mm. It should be pointed out that the grinding and packing processes were carried out in a glove box filled with high purity argon. The samples were divided into six groups according to the K-doping level. The samples were first sintered at 500 °C for 15 h and then followed with a second sintering at 900 °C for 35 hours. In order to prevent sample oxidation, high purity argon gas was flowing surrounding the samples during the sintering processes.

The phase identification was characterized using X-ray diffraction (XRD) with Cu-Kα radiation. The diffraction peaks could be well indexed on the basis of tetragonal $ThCr_2Si_2$-type structure with the space group I4/mmm. The superconducting properties were studied by magnetization and standard four-probe resistivity measurements using a physical property measurement system (Quantum Design). The critical current density $J_c$ was determined using the Bean model. Microstructure was measured by scanning electron microscopy (SEM) and transmission electron microcopy (TEM) after peeling away the Ta sheath.

Figure 1 shows the X-ray diffraction patterns of $Ba_{0.6}K_{0.4+x}Fe_2As_2$ superconductors with $x$ ranging from 0 to 0.1 in step of 0.02. All samples exhibit $ThCr_2Si_2$-type structure with no detectable impurity phases. Lattice parameters are not strong dependent of potassium content as indicated by data shown in Table 1. Note that SEM/EDX analysis confirms that there is always a slight K-loss for all the samples with different K-doping levels because of high sintering temperature Thus, there is a small increase (decrease) in $c$- ($a$-) axis lattice parameter with increasing K-doping level from $x = 0$ to $x = 0.02$. However, the $a$- and $c$-axis lattice parameters nearly keep as a constant with further increase of K-doping level, indicating the Ba loss is not obvious and element K can not substitute Ba.

Figure 2 shows resistivity versus temperature curves ($\rho$-$T$) for samples with different K-doping levels. All the samples exhibit metallic characteristic before superconducting transition with sharp superconducting transition. For the samples of $x \leq 0.06$, the onset $T_c$ = 37 $K$ and $\Delta T_c \approx$ 2 $K$ were measured. With further increase in K-doping level, $T_c$ = 36.5 $K$ and $\Delta T_c \approx$ 3 $K$ were measured. Large residual resistivity ratio (*RRR*) of $\approx$ 7 was determined for the samples with $x \leq 0.06$, and the *RRR* decreases to about 4 for the samples with increasing K-doping level. According to metallic material Matthiessen's rule, the slope of $\alpha = d\rho(T)/dT$ reflects the temperature dependent lattice scattering. Decrease in residual resistivity ratio ($R$(300 $K$)/$R$(38 $K$)) and increase in lattice scattering coefficient ($\alpha$) indicate the enhancement of lattice scattering.

The magnetization hysteresis curves are measured at 5 $K$ for samples of various doping level (0 $\leq$ $x$ $\leq$ 0.1). Only these curves measured on $Ba_{0.6}K_{0.4}Fe_2As_2$ and $Ba_{0.6}K_{0.5}Fe_2As_2$ samples are shown in fig. 2 (inset) as examples. Curves measured on others samples are similar in shape. Symmetric feature of the curves indicates that the bulk current rather than surface shielding current is the predominant under superconducting state. The gap in the loop drops drastically in the low-field region with increasing applied field but it rapidly becomes very smooth over a wide range of applied magnetic field. According to the hysteretic curve, nearly no ferromagnetic background can be detected, indicating good quality for our samples.

Figure 3 illustrates the magnetic field dependence of the critical current density $J_c$ derived from the hysteresis loop width using the extended Bean model of $J_c = 20\Delta M / a(1 - a/3b)$ taking the full sample dimensions. Where $\Delta M$ is the width of magnetization loop, $a$ and $b$ denote the dimensions of the sample perpendicular to the direction of magnetic field ($a$ < $b$). As shown in figure 3, the $J_c$ shows strong dependence on K-doping level. $J_c$ increases with increasing K content and a maximum was observed for sample with $x$ = 0.1 in present study. Attempt of further increase in K content led to sample instability in air.

We observed high overall magnetic $J_c$ in the samples with K over-doping. It is known that two kinds of loops, intra-grain and inter-grain, can contribute to the magnetization of granular superconductors. To separate these two, we measured magnetization hysteresis loops on powder samples of $Ba_{0.6}K_{0.4}Fe_2As_2$ and $Ba_{0.6}K_{0.5}Fe_2As_2$. If we assume the current flows only within the grains ($J_c = 30\Delta M / R$, R is the average grain size), the $J_c$ based on the individual grains (the average grain size is about 10 $\mu m$) would illuminate the intra-grain current loops. According to the present experiments, the $J_c$ based on the individual grains are obviously enhanced by over-doping of K, as shown in the inset of Fig. 3. Thus, it is reasonable to infer that the increase of $J_c$ in K over-doped samples can be attributed to the enhancement of intra-grain currents.

In order to understand the mechanism for improved intra-grain currents in K over-doped samples, we performed SEM and TEM studies. Fig. 4 (a) and (b) illustrate the typical SEM images for samples with $x = 0$ and 0.1, respectively. Microstructures measured by SEM for other samples are very similar. As the SEM study on microstructures reveals, there is no clear difference in grain size for all the samples. Thus, the enhanced $J_c$ can not be due to the growth of grain size. Fig.4 (c) and (d) show the high-resolution TEM micrographs of the $Ba_{0.6}K_{0.4}Fe_2As_2$ and $Ba_{0.6}K_{0.5}Fe_2As_2$ samples, respectively. Compared with $Ba_{0.6}K_{0.4}Fe_2As_2$, higher concentration of dislocations is detected in $Ba_{0.6}K_{0.5}Fe_2As_2$ samples, as shown in fig. 4(d). This is confirmed by *R-T* data, as shown in fig. 2. Normal state resistance ($\rho$) partially reflects the strength of lattice scatting ($\beta T^5$, $\beta$ is a constant). The marked increase in normal state resistances for the samples with higher K-doping level indicate the enhancement of lattice scatting, which is consistent with the TEM observation. It should be pointed out that numerous of randomly selected grains have been examined by TEM, confirming the fact that there is much higher concentration of dislocations in $Ba_{0.6}K_{0.5}Fe_2As_2$ than that in $Ba_{0.6}K_{0.4}Fe_2As_2$. Small distortions of lattice induced by over-doped K led to lattice dislocations, resulting in enhanced pining force and hence improved critical current density $J_c$. Our results suggest that it is beneficial to prepare $Ba_{0.6}K_{0.4}Fe_2As_2$ samples with a slight excess K for high $J_c$ applications.

In summary, the $Ba_{0.6}K_{0.4}Fe_2As_2$ samples prepared with a slight excess of K demonstrate only about 0.5 $K$ degradations in $T_c$ and greatly enhancement of in–field $J_c$. There are no obvious differences in grain size and connectivity between grains. The superior $J_c$ in K-superfluous $Ba_{0.6}K_{0.4}Fe_2As_2$ samples is attributed to the enhanced pining force originated from the increase in dislocations. Our results show that it is indispensable to prepare the $Ba_{0.6}K_{0.4}Fe_2As_2$ samples with appropriate K excess for obtaining excellent superconducting properties.

The authors thank Beihai Ma at Argonne national Laboratory (ANL, USA) for his help in English corrections. This work is partially supported by the Beijing Municipal Science and Technology Commission under Grant No. Z09010300820907, National '973' Program (Grant No. 2011CBA00105) and the National Natural Science Foundation of China (Grant No. 51025726).

**Table**

Table 1. Lattice constants for $Ba_{0.6}K_{0.4+x}Fe_2As_2$ samples with x ranging from 0 to 0.1.

| Composition | Lattice parameters | |
|---|---|---|
| | $a$ (Å) | $c$ (Å) |
| $Ba_{0.6}K_{0.40}Fe_2As_2$ | 3.9127 | 13.3177 |
| $Ba_{0.6}K_{0.42}Fe_2As_2$ | 3.9096 | 13.3362 |
| $Ba_{0.6}K_{0.44}Fe_2As_2$ | 3.9095 | 13.3355 |
| $Ba_{0.6}K_{0.46}Fe_2As_2$ | 3.9103 | 13.3378 |
| $Ba_{0.6}K_{0.48}Fe_2As_2$ | 3.9115 | 13.3362 |
| $Ba_{0.6}K_{0.50}Fe_2As_2$ | 3.9095 | 13.3589 |

**Figures**

Figure 1. XRD patterns of $Ba_{0.6}K_{0.4+x}Fe_2As_2$ samples with x ranging from 0 to 0.1 in step of 0.02.

Figure 2. Resistivity versus temperature curves (ρ-T) for $Ba_{0.6}K_{0.4+x}Fe_2As_2$ samples with different K-doping levels. Inset: Magnetization hysteretic loops of the samples with x = 0 and x = 0.1 at 5K.

Figure 3. Magnetic field dependences of critical current densities at 5 K for bulk samples with various K-doping levels. Inset: Magnetic field dependences of $J_c$ at 5 $K$ for powder samples with x = 0 and x = 0.1..

Figure 4. Upper: SEM micrographs for the samples with (a) x = 0 and (b) x = 0.1. Lower: TEM images for the samples with (c) x = 0 and (d) x = 0.1.

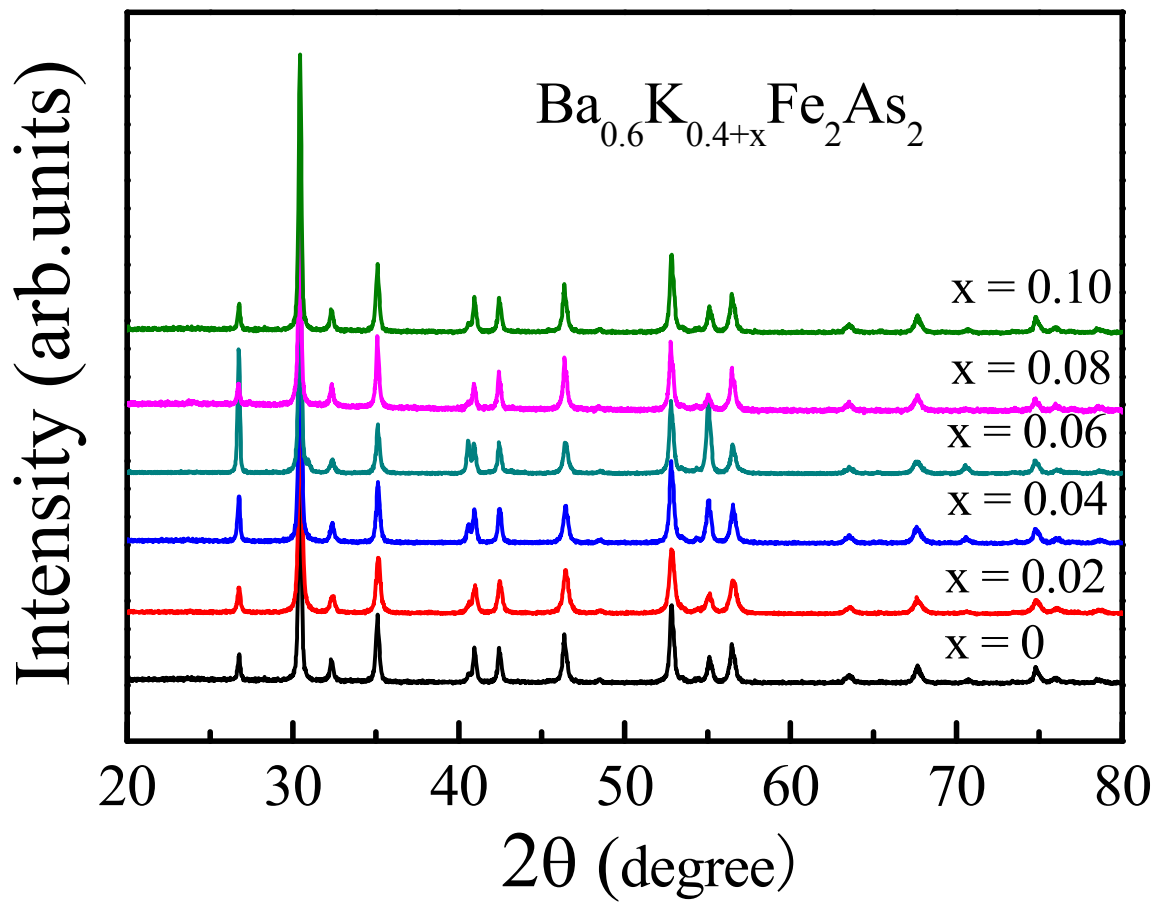

Fig. 1 Wang et al

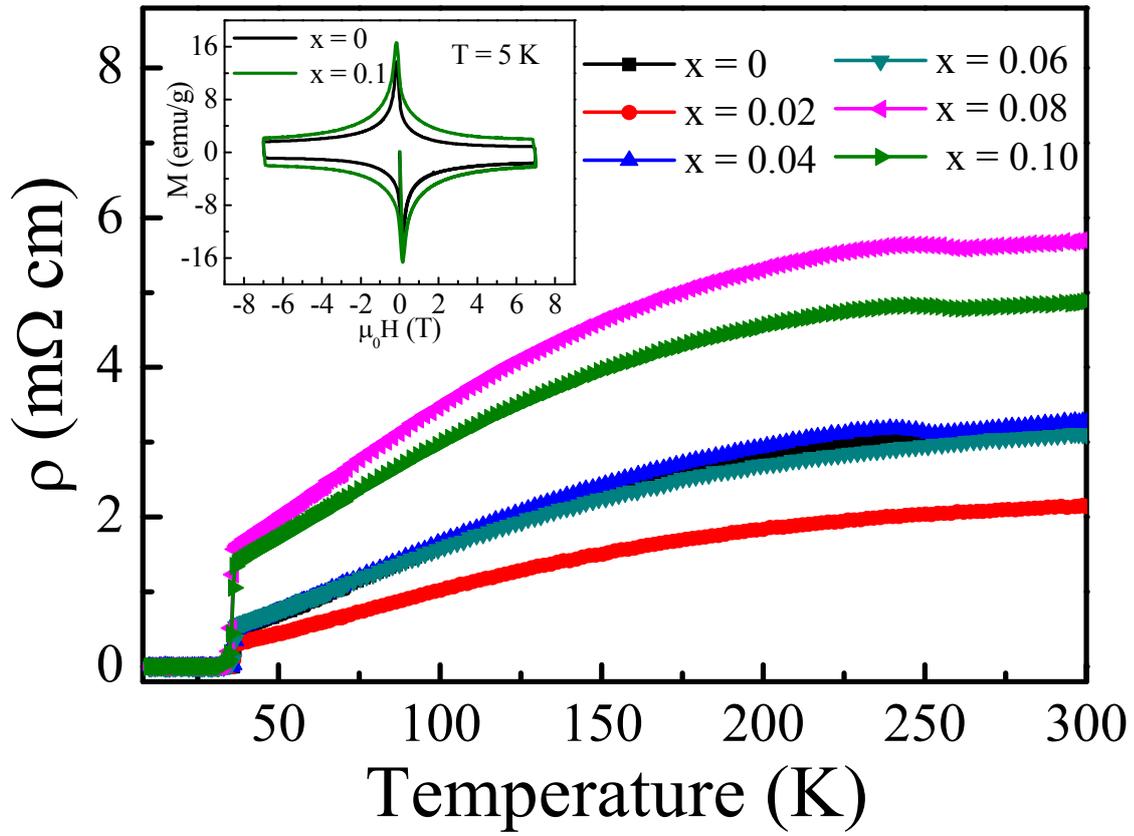

Fig. 2 Wang et al

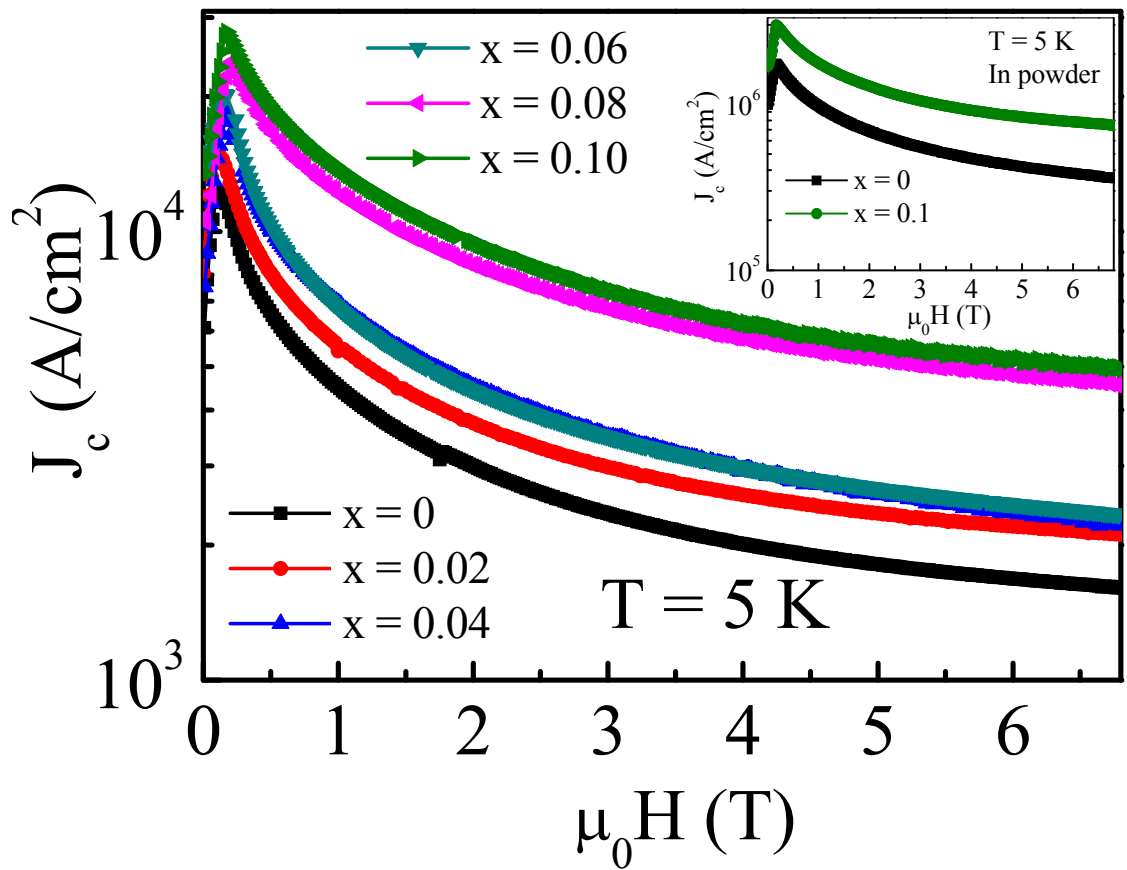

Fig. 3 Wang et al

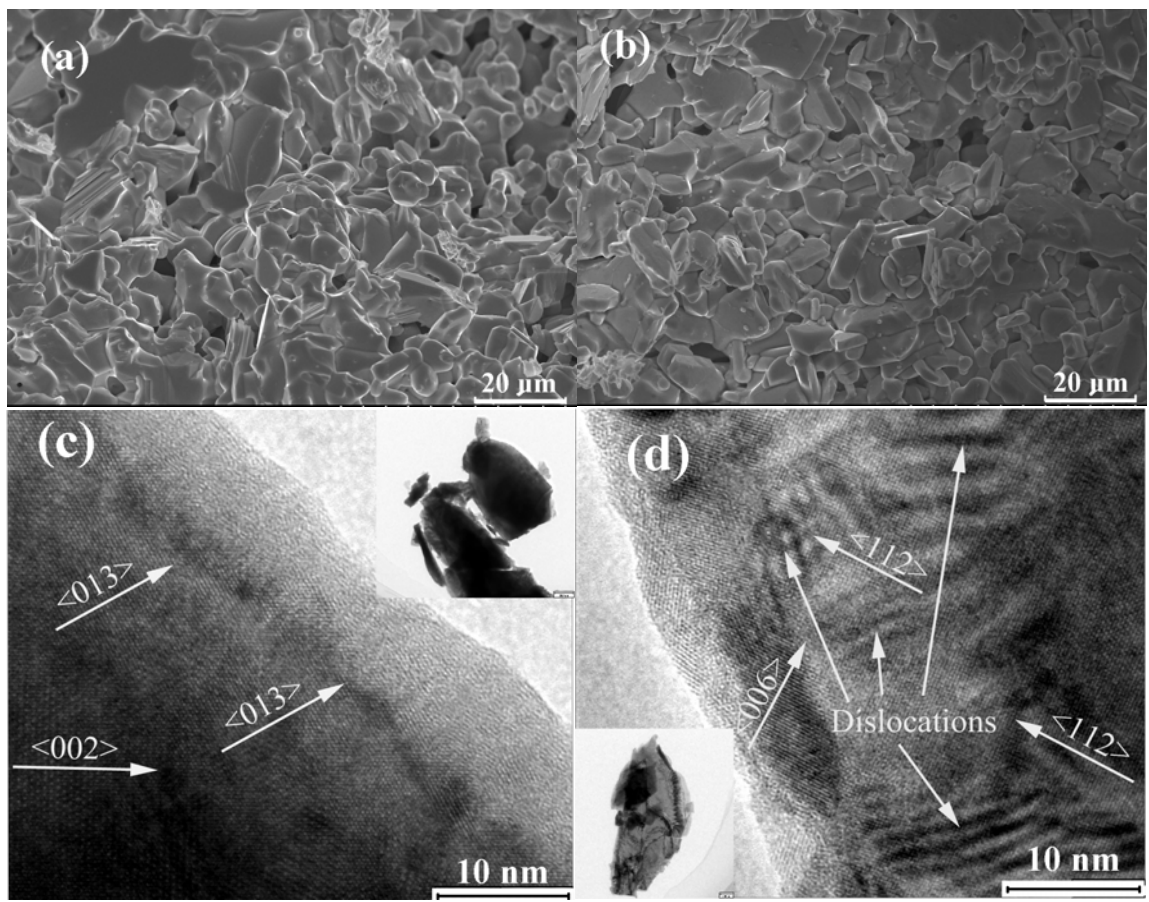

Fig. 4 Wang et al